\documentclass[preprint,A4]{elsarticle}
\usepackage[margin=2cm]{geometry}
\usepackage[hidelinks]{hyperref}

\usepackage{amssymb}
\usepackage{amsmath}
\journal{Statistics \& Probability Letters}

\begin{document}

\begin{frontmatter}



\title{Bandwidth of Gamma-Distribution-Shaped Functions 
via Lambert W Function}


\author{Anthony LoPrete${}^{a}$} %% Author name
\author{Johannes Burge${}^{a,b,c}$}

\affiliation{organization={Bioengineering Graduate Group, ${}^b$Neuroscience Graduate Group, ${}^c$Department of Psychology, 
\\ University of Pennsylvania},
            city={Philadelphia},
            state={Pennsylvania},
            country={USA}}

%% Abstract
\begin{abstract}
The full width at half maximum (FWHM) is a useful quantity for characterizing the bandwidth of unimodal functions. However, a closed-form expression for the FWHM of gamma-shaped functions—i.e. functions that are shaped like the gamma distribution probability density function (PDF)—is not widely available. Here, we derive and present just such an expression. To do so, we use the Lambert W function to compute the inverse of the gamma PDF. We use this inverse to derive an exact analytic expression for the width of a gamma distribution at an arbitrary proportion of the maximum, from which the FWHM follows trivially. (An expression for the octave bandwidth of gamma-shaped functions is also provided.) The FWHM is then compared to the Gaussian approximation of gamma-shaped functions. A few other related issues are discussed.
\end{abstract}

%% Keywords

\end{frontmatter}



\section{Introduction}
\label{introduction}
%% Labels are used to cross-reference an item using \ref command.

Functions that are shaped like the gamma distribution probability density functions (PDFs) are widely used across science, engineering, and business. In vision research and systems neuroscience, gamma-shaped functions are used to characterize the temporal response properties of photoreceptors to photon absorptions \cite{CAO2007}, the distribution of contrast energy across images of natural scenes \cite{IYER2019}, and the characterization of cross-correlograms obtained from continuous psychophysics experiments \cite{CHIN2022,BURGE2025}. In archaeology and other fields that make use of radioisotope decay, gamma distributions are useful for radiocarbon dating \cite{BLAAUW2011}. In electrical engineering and digital signal processing, gamma distributions make frequent appearances – including when modeling the signal-to-noise ratio of wireless channels and verifying vehicles with Gabor filters \cite{ATAPATTU2011,GUO2014}. In operations research, gamma distributions are used in queuing theory to describe the distribution of wait times \cite{ALLEN1978}. In Bayesian statistics, the gamma distribution is the conjugate prior for Poisson likelihood functions \cite{GELMAN1995}. Its widespread use is partly due to the fact that the gamma distribution is a super-family distribution that contains the exponential, chi-squared distribution, and Erlang distributions as special cases. 

Gamma-distribution-shaped functions have the shape of the gamma distribution PDF but do not necessarily integrate to $1.0$ and can have their domains shifted. Specifically, gamma-distribution-shaped functions are given by

\begin{equation}
\label{eq:gamshape}
g(x)=K x^{\prime a-1}\exp{\left(-x^{\prime}/b\right)}
\end{equation}
where $a>0$ is the shape parameter, $b>0$ is the scale parameter, $\exp$ is the exponential function, $K$ is an arbitrary scalar, and $x^\prime=x+s$  where $s$ is an arbitrary shift to the domain. Gamma-distribution-shaped functions can be converted to gamma distribution PDFs by setting $K$ to an appropriate scalar and $s=0$ (see Eq. \ref{eq:gampdf}).

\begin{figure}[t]
\centering
\includegraphics[width=0.65\textwidth]{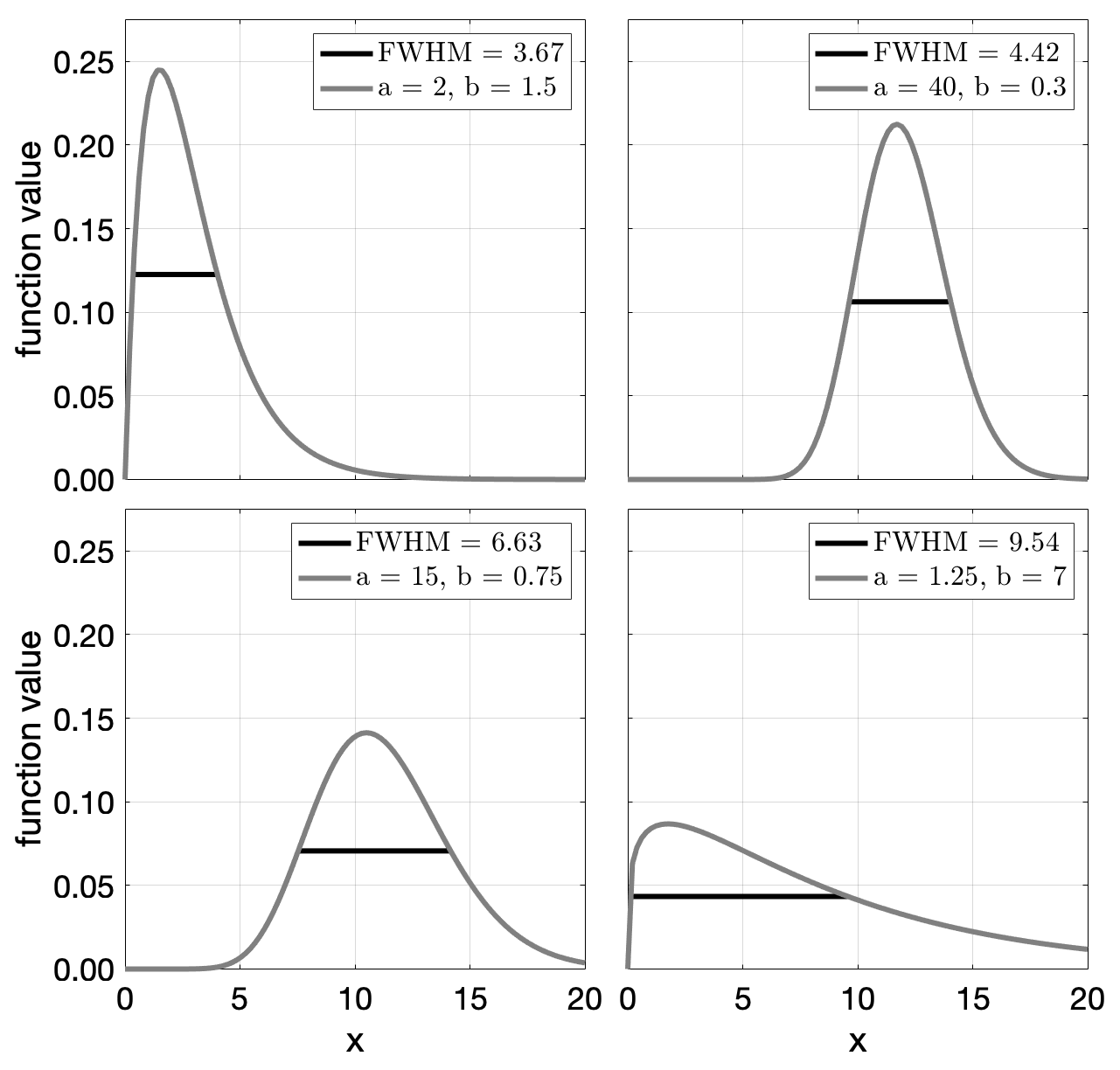}
\caption{Example gamma probability densities (i.e. gamma-shaped functions; gray curves) and their respective FWHMs (black lines). Values for the FWHM, and the shape and scale parameters are indicated in the legend.}
\label{fig:gamfwhm}
%% https://en.wikibooks.org/wiki/LaTeX/Importing_Graphics#Importing_external_graphics
\end{figure}

Empirical scientists often consider it useful to describe a function by its spread. The full width at half maximum (FWHM) is a common such descriptor. The FWHM is defined as the distance between arguments corresponding to half of the function's maximum (Fig. \ref{fig:gamfwhm}). To find the exact value for the FWHM of a unimodal function, one must find an expression for the function inverse—or more precisely, the converse relation—, evaluate it at half the function’s maximum value, and then subtract the lesser from the greater output.  However, an exact closed-form expression for the FWHM of gamma-distribution-shaped function is not readily available in the literature.

Here, we derive a closed-form expression for the FWyM--that is, the full width at an arbitrary proportion of the maximum–of gamma-distribution-shaped functions, from which an expression for the FWHM trivially follows. 

For gamma-distribution-shaped functions, the FWHM—and, more generally, the FWyM—is defined only when the shape parameter $a$ is greater than or equal to one. When the shape parameter $a$ is less than one, the gamma distribution does not have a properly defined mode; the consequence is that the half-maximum—or any other proportion of the maximum—is not well-defined. For shape parameter $a = 1$, the gamma distribution corresponds to the exponential distribution, for which the FWHM can be easily determined via algebraic methods (i.e. $\textrm{FWHM}_{\textrm{exp}} =b \ln{2}$ ). This article, therefore, focuses on gamma-distribution-shaped functions having shape parameter $a > 1$. 

The main result of the paper is as follows: the FWHM for all gamma-distribution-shaped functions with shape parameter $a > 1$ and scale parameter $b$ is given by

\begin{equation}
\label{eq:gamfwhm}
\textrm{FWHM} = (a-1)b \left( W_{0} \left(-\left(1/2\right)^{1/(a-1)}/e\right) - W_{-1} \left(-\left(1/2\right)^{1/(a-1)}/e \right) \right)
\end{equation}
where $W_k$ is the kth  branch of the Lambert W function. \footnote{There are infinitely many branches of the Lambert W function—each indexed by a unique integer $k$—and each inverting expressions of the form $u \exp{(u)} $. When considering real-valued inputs and outputs, only the $k = 0$ and $k = -1$ branches are relevant, as the other branches always return complex numbers.} The Lambert W function inverts expressions of the form $u\exp{(u)}$, such that $W_k(u\exp{(u)})=u$ (Fig. 2). Such an expression appears when deriving the FWHM for gamma-distribution-shaped functions (see Results). The Lambert W function thus makes finding the FWHM of a gamma-shaped function straightforward \cite{VEBERIC2012}. However, the Lambert W function is not as well-known as it perhaps ought to be \cite{KESISOGLOU2021}, which may be why an analytical expression of the gamma FWHM is not readily available in the literature. Implementations of the Lambert W function are available in many programming languages and computer algebra systems (e.g. MATLAB, Mathematica), and hence makes it simple to compute the FWHM of gamma-distribution-shaped functions using modern scientific computing languages (see Section \ref{sec:matlab}).

In the remainder the paper, we derive the gamma distribution's inverse PDF and use it to write an expression of the FWyM from which the FWHM follows trivially. We compare this analytic FWHM to a Gaussian approximation of the gamma distribution's FWHM. We also present an analytic expression for the octave bandwidth of gamma-distribution-shaped functions. Lastly, we briefly discuss a provided MATLAB function implementing the results of this paper.

\section{Results}
\label{results}
This section presents the mathematical results of the paper. The preliminary details—the definition of gamma-distribution-shaped functions, the definition of the FWHM, and the Lambert W function—are presented in the introduction. Our derivations use the gamma distribution PDF because it is the most common gamma-distribution-shaped functions function (see Eq. \ref{eq:gamshape}), and because arbitrary scalings and shifts to the gamma distribution PDF do not change the results.

If a random variable is gamma distributed—that is, $X \sim \textrm{Gamma}(a,b)$—with support $x \geq 0$, then its probability density $p(x)$ is given by 

\begin{equation}
\label{eq:gampdf}
p\left(x\right)=\left(\frac{1}{\Gamma\left(a\right)b^a}\right)x^{a-1}\exp{\left(-x/b\right)}
\end{equation}
where $\Gamma\left(\bullet\right)$ is the gamma function. The gamma function is a generalization of the factorial function $\Gamma\left(x\right)=\left(x-1\right)!$ such that for non-integer values of x, the gamma function smoothly interpolates between the outputs of the integer-valued factorial function. 

%% Use \subsection commands to start a subsection.
\subsection{Gamma Probability Density and its Inverse}
\label{inverse}

Here, we derive the inverse (i.e. converse relation) of the gamma probability density function. For notational simplicity, we denote the probability $p\left(x\right)$ as $p$. 

Multiplying both sides of Eq. \ref{eq:gampdf} by $\Gamma(a)b^a$ and raising both sides to the power $1/(a - 1)$ yields

\begin{equation}
\label{eq:multexp}
\left(p\Gamma\left(a\right)b^a\right)^{1/\left(a-1\right)}=x\exp{\left(\frac{-x}{(a-1)b}\right)}
\end{equation}
Multiplying both sides by $-1/((a - 1)b)$ gives

\begin{equation}
\label{eq:mult}
\frac{-\left(p \Gamma (a) b^a \right)^{1/(a-1)}}{(a-1)b} = \frac{-x}{(a-1)b} \exp{\left(\frac{-x}{(a-1)b}\right)}
\end{equation}
Observing that there is an expression of the form $u\exp{u}$ on the right-hand-side of Eq. \ref{eq:mult}, we substitute $u=-x/((a-1)b)$ to obtain

\begin{equation}
\label{eq:usub}
\frac{-\left(p \Gamma (a) b^a \right)^{1/(a-1)}}{(a-1)b} = u \exp{(u)}
\end{equation}
Applying the Lambert W function to both sides

\begin{equation}
\label{eq:lambertw}
W_k \left(\frac{-\left(p \Gamma (a) b^a \right)^{1/(a-1)} }{(a-1)b} \right) = W_k \left( u \exp{(u)} \right)
\end{equation}
Simplifying the right-hand side 

\begin{equation}
\label{eq:invsimplify}
W_k \left(\frac{ -\left(p \Gamma (a) b^a \right)^{1/(a-1)} }{(a-1)b} \right)  = u
\end{equation}
Resubstituting $-x/((a-1)b)=u$ gives

\begin{equation}
\label{eq:resubstitute}
W_k \left(\frac{ -\left(p \Gamma (a) b^a \right)^{1/(a-1)} }{(a-1)b} \right)  = \frac{-x}{(a-1)b}
\end{equation}
Finally, multiplying through by $b(a-1)$  and substituting $f_k^{-1}(p)$ for $x$ gives:

\begin{equation}
\label{eq:gampdfinv}
f_{k}^{-1}(p) = -(a-1)b W_k \left(\frac{-\left(p \Gamma (a) b^a \right)^{1/(a-1)} }{(a-1)b} \right)
\end{equation}
Hence, Eq. \ref{eq:gampdfinv} is the inverse (i.e. converse relation) of the gamma probability density function. 

\begin{figure}[t]
\centering
\includegraphics[width=0.75\textwidth]{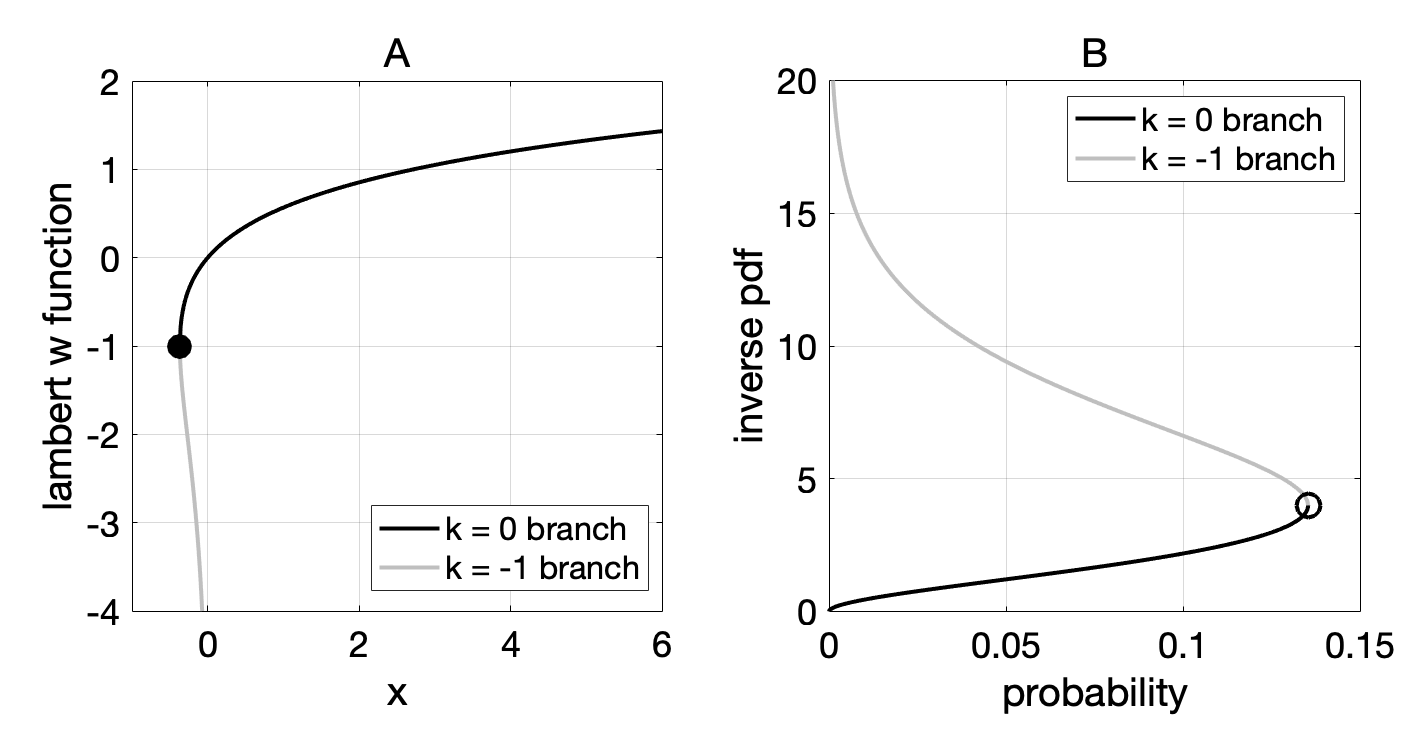}
\caption{\textbf{A} Real-valued branches of the Lambert W function for the $k = 0$ branch (black) and $k=-1$ branch (gray). Both the $k=-1$ and $k=0$ branches of the Lambert W function equal $-1$ when evaluated at $x = -1/e$ (black dot). \textbf{B} The value of the inverse (i.e. converse relation) of the Gamma(3,2) probability density evaluated using the $k=0$ and $k = -1$ branches of the Lambert W function (black and gray, respectively). Note that the $k = -1$ branch returns values above the mode (white dot), whereas the $k = 0$ branch returns values below the mode.}
\label{fig:lambertw}
\end{figure}

Note that because $W_k(\bullet)$ is a multi-valued function—one value for each branch k—the converse relation $f_k^{-1}(p)$ is a multivalued function. Recall that our aim is to obtain the arguments on each side of the mode of the density function that correspond to the desired proportion of the function maximum. For this purpose, the relevant branches of the Lambert W function are indexed at $k = 0$ and $k = -1$ (Fig. \ref{fig:lambertw}A). Evaluating Eq. \ref{eq:gampdfinv} for the $k = 0$ branch of the Lambert W function returns the argument below the mode, while evaluating at the $k = -1$ branch returns the argument above the mode (Fig. \ref{fig:lambertw}B). When evaluated at the mode itself, the output is equal for both branches. \footnote{Equation \ref{eq:gampdfinv} bears an interesting resemblance to the quantile function of the $\textrm{Gamma}(2,b)$ distribution (see \ref{app:quantile}), in that they are both written in terms of the $k=-1$ branch of the Lambert W function. The median can be computed from the quantile function, and studies of the median of the gamma distribution PDF have important theoretical implications in a number of fields of mathematics.}

\subsection{Full Width of Functions (FWyM and FWHM)}
The Full Width at y-Max (FWyM) of a unimodal function is the full width of the function at an arbitrary proportion of the maximum \cite{SALVI2023,VESTERGAARD2000}. The FWHM is the distance between half-maxima on each side of the mode.  To find the FWyM, we evaluate the converse relation (Eq. \ref{eq:gampdfinv}) at the $k = -1$ and $k=0$ branches with a value that is the desired proportion of the function maximum, and then subtract the smaller from the larger value.

The mode of the gamma probability density is given by $(a-1)b$. The desired proportion y of the maximum $p((a-1)b)$ is given by

\begin{equation}
\label{eq:ymaxprop}
y*p((a-1)b) = \frac{y}{\Gamma(a)b^a}((a-1)b)^{a-1}e^{-((a-1)b)/b}
\end{equation}
Substituting Eq. \ref{eq:ymaxprop} into Eq. \ref{eq:gampdfinv} yields

\begin{equation}
x = -(a-1)b W_k \left(\frac{1}{(a-1)b}\left(-\frac{y}{\Gamma(a)b^a}((a-1)b)^{a-1}e^{-((a-1)b)/b} \Gamma (a) b^a \right)^{1/(a-1)}\right)
\end{equation}
Simplifying

\begin{equation}
\label{eq:fwymsimplify}
x = -(a-1)b W_k \left(-y^{1/(a-1)}/e \right)
\end{equation}
The FWyM is the difference between the two values of $x$ obtained from this equation corresponding to the two relevant branches of the Lambert W function. The $k = -1$ branch returns a larger value than the $k = 0$ branch on our domain. Hence, the FWyM is given by

\begin{equation}
\label{eq:gamfwymunsimp}
\textrm{FWyM} = \left(-(a-1)b W_{-1} \left(-y^{1/(a-1)}/e \right) \right) - \left(-(a-1)b W_{0} \left(-y^{1/(a-1)}/e \right) \right) 
\end{equation}
Simplifying and re-arranging yields

\begin{equation}
\label{eq:gamfwym}
\textrm{FWyM} = (a-1)b \left( W_{0} \left(-y^{1/(a-1)}/e \right) - W_{-1} \left(-y^{1/(a-1)}/e \right) \right)
\end{equation}
where $y$ can be set to any positive value between $0$ and $1$ to obtain the full width of the gamma density at any arbitrary proportion of the maximum (see Fig. 3). This expression for the FWyM of the gamma probability density holds for all gamma-distribution-shaped functions. As noted, the most widely used version of this expression is the full width at half maximum (FWHM), which is obtained by simply setting $y = 1/2$ (Eq. \ref{eq:gamfwhm}).

\begin{figure}[t]
\centering
\includegraphics[width=0.4\textwidth]{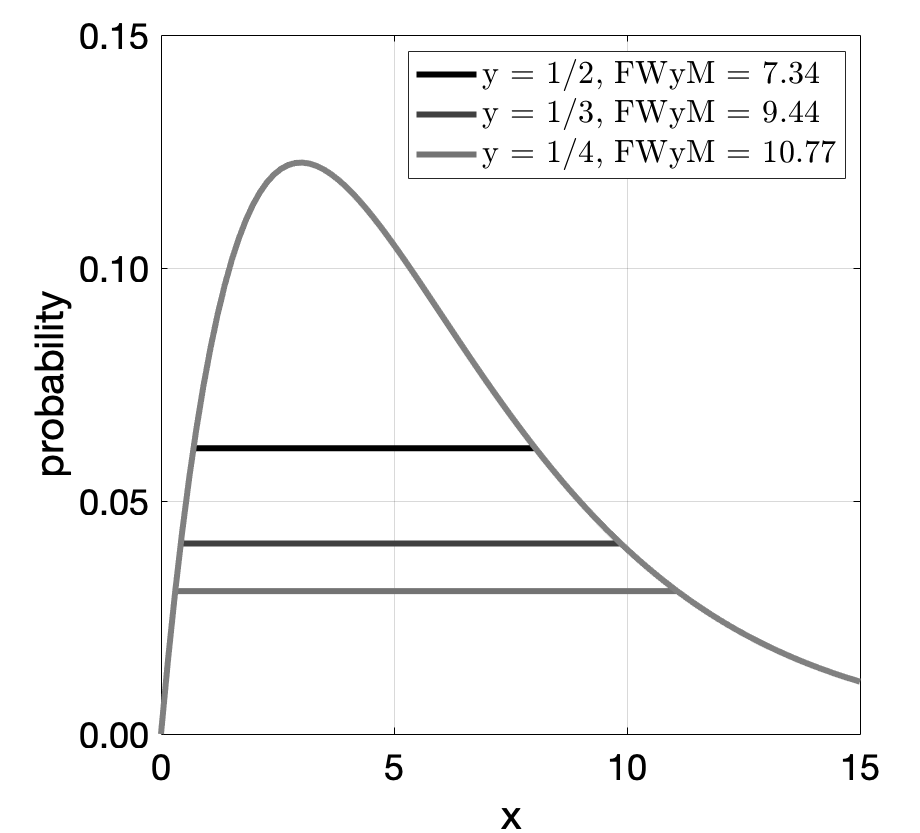}
\caption{FWyMs of the Gamma(2,3) distribution for three different values of y (i.e. $1/2$, $1/3$, $1/4$). }
\label{fig:gamfwym}
\end{figure}

\subsection{Comparison to Gaussian Approximation}

The gamma distribution is well-approximated as a Gaussian distribution for large values of the shape parameter $a$. The central limit theorem holds that the sum of $n$ independently identically distributed (i.i.d.) random variables $X_1$, $X_2$, $\dots$, $X_n$  tends toward Gaussian
\begin{equation}
\sum_{i=1}^{n} X_i \ \dot{\thicksim} \ \mathcal{N}(n \mu,n\sigma^2)
\end{equation}
where each random variable $X_i$  has the same mean $\mu$ and variance $\sigma^2$.

The $\textrm{Gamma}(a,b)$ distribution can be described as the sum of $a$ i.i.d exponentially distributed random variables with mean $b$

\begin{equation}
\label{eq:gammadistexp}
\textrm{Gamma(a,b)} \thicksim \sum_{i=1}^{a} X_i \quad \textrm{where} \quad X_i \thicksim \textrm{exp(b)}
\end{equation}
The mean (i.e. expected value) and variance of of each random variable $X_i$ equal  $\mu_i  = b$ and $\sigma_i^2 = b^2$, respectively. 
Hence, from the central limit theorem, the random variable $X \sim \textrm{Gamma}(a,b)$ is approximately distributed as:

\begin{equation}
X \phantom{a} \dot{\sim} \phantom{a} \mathcal{N}\left(ab,ab^2\right)
\end{equation}
Applying the standard definition for the FWHM of a Gaussian distribution of $\textrm{FWHM} = 2\sqrt{2\log{2}}\sigma$  with $\sigma\ = b\sqrt{a}$, yields the Gaussian approximation to the gamma distribution FWHM

\begin{equation}
\textrm{FWHM} \approx\left(2\sqrt{2\log{2}}\right)b\sqrt a
\end{equation}
where $ab^2$ is the variance of the gamma-distributed random variable.

\begin{figure}[t]
\centering
\includegraphics[width=\textwidth]{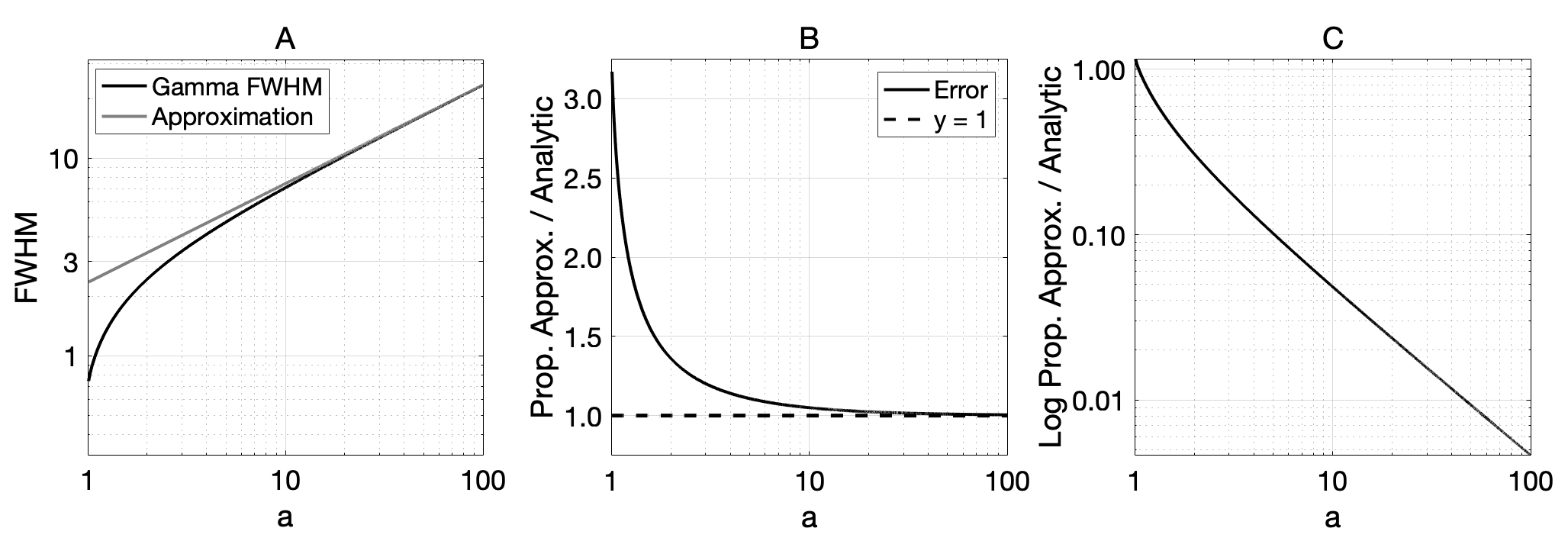}
\caption{Gamma FWHM versus its Gaussian-approximation. \textbf{A} FWHM as a function of the shape parameter $a$, for gamma distributions with scale parameter $b = 1$. \textbf{B} Proportional error between the Gaussian-approximation and true value of the FWHM. Note that the proportional error function is invariant to the value of the scale parameter b. \textbf{C} Logarithm of the proportional error between Gaussian-approximate and analytical gamma FWHM plotted against $a$ on a log-x log-y plot.}
\label{fig:fwhmapprox}
\end{figure}

This is a good approximation, particularly for large values of a, and is not as complicated as the analytical FWHM expression. However, it always overshoots the true value. Figure \ref{fig:fwhmapprox} compares the difference between the analytical and CLT-approximate gamma FWHM and shows the proportion by which the approximation exceeds the true FWHM, indicating that this ratio asymptotically approaches $1$ as $a$ increases. 

\subsection{Octave Bandwidth}

The octave bandwidth is another frequently used characterization of a function’s width, that is closely related to the bandwidth (or FWHM) of a function \cite{IYER2019}. Octave bandwidth is given by

\begin{equation}
\textrm{BW}_{\textrm{oct}}=\log_2\left(\frac{H}{L}\right)
\end{equation}

where $H$ and $L$ denote the high and low FWHM components, respectively. Given expressions for the inverse of probability density function (see. Eq. \ref{eq:gampdfinv}) and the mode, obtaining an expression for the octave bandwidth is easy.

Here, we provide the expression for the octave bandwidth of gamma-distribution-shaped functions. As mentioned previously, when deriving the FWyM and FWHM, we found that $H$ corresponds to the $k = -1$ branch and that $L$ corresponds to the $k = 0$ branch of the Lambert W function. Substituting 

\begin{equation}
\textrm{BW}_{\textrm{oct}} = \log_2\left( \frac{-(a-1)b W_{-1} \left(\frac{-(\frac{1}{2})^{1/(a-1)}}{e} \right)}{-(a-1)b W_{0} \left(\frac{-(\frac{1}{2})^{1/(a-1)}}{e} \right)} \right)
\end{equation}
Simplifying

\begin{equation}
\label{eq:bwoct}
\textrm{BW}_{\textrm{oct}} = \log_2\left( \frac{W_{-1} \left(\frac{-\left(\frac{1}{2}\right)^{1/(a-1)}}{e} \right)}{W_{0} \left(\frac{-\left(\frac{1}{2}\right)^{1/(a-1)}}{e} \right)} \right)
\end{equation}
Equation \ref{eq:bwoct} has appeared previously in the literature, but has not been described as the octave bandwidth of a gamma-distribution-shaped function. It has been reported as the frequency domain octave bandwidth of a Cauchy filter \cite{BOUKERROUI2004}. The Cauchy filter in Fourier domain takes the form of a gamma distribution, which may be why this equation appears without any explicit mention of the gamma distribution. 

Equation \ref{eq:bwoct} specifies the octave bandwidth—the log-base-2 ratio of the high to the low values—spanning the full width at half maximum of the function. As is the case with the full width, the octave bandwidth can be generalized to an arbitrary proportion y of the maximum value of the function. For gamma-shaped functions, the generalized octave bandwidth is given by

\begin{equation}
\label{eq:bwocty}
\textrm{BW}_{\textrm{oct}} = \log_2\left( \frac{W_{-1} \left(\frac{-y^{1/(a-1)}}{e} \right)}{W_{0} \left(\frac{-y^{1/(a-1)}}{e} \right)} \right)
\end{equation}

\subsection{MATLAB Function}
\label{sec:matlab}

MATLAB functions are provided on the MATLAB file exchange (see \url{https://www.mathworks.com/matlabcentral/fileexchange/181960-gamma-distribution-bandwidth}), which compute the FWyM (and, hence, FWHM) and octave bandwidth of gamma distributions using the formulae derived in this paper. Both functions depend on MATLAB’s Symbolic Math Toolbox, which includes an implementation of the Lambert W function. 

\section{Discussion and Conclusion}
\label{discussion}
Many kinds of data and signals are well characterized by functions taking their shape from gamma distribution PDFs. The full width at half-maximum (FWHM) is a common measure of the bandwidth of unimodal functions. This article derived a general expression for the FWHM of gamma-distribution-shaped functions for all cases in which the FWHM is well-defined ($a>1$, $b>0$). We derived the multivalued inverse (i.e. converse relation) of the gamma probability density function and used it to derive an expression for the FWyM from which the FWHM trivially follows.

For use in scientific applications, a simple MATLAB function was also provided.

Future work may consider investigating the bandwidth of the generalized gamma distribution a super-family distribution, which contains the gamma probability density as a special case \cite{STACY1962}. Deriving the bandwidth of generalized gamma distributions may be impossible outside the special case of the gamma distribution. If so, a proof showing so could be valuable.

\section*{Acknowledgments}
This work was supported by the National Eye Institute and the Office of Behavioral and Social Sciences Research–National Institutes of Health Grant R01-EY028571 to J.B.

\appendix

\section{Special-Case Equivalence of Quantile and Inverse Density Functions}
\label{app:quantile}

The gamma distribution's median, $\upsilon(a,b)$, is the value at which the gamma distribution's cumulative distribution function (CDF, notated $F(x)$) equals one half 

\begin{equation}
 F\left(\nu (a,b)\right)=\frac{1}{2} \phantom{-} \textrm{where} \phantom{`} \phantom{`} F(x)= \int_{0}^{x}\frac{1}{\Gamma(a)b^a}x^{a-1}e^{-x/b}dx
\end{equation}
Neither the median nor the CDF of the gamma distribution have known closed-form expressions in terms of elementary functions for all values of the shape parameter \footnote{The scale parameter's only influence on the gamma distribution’s median is a scaling of the form $\upsilon(a,b)=b\ast\upsilon(a,1)$.}. It is useful to have the important distributional characteristics (median, mean, variance, etc.) for probability distributions well-understood if not defined exactly, rendering the gamma distribution's median an active area of research. The median is known to be bounded by $a-1/3<\upsilon(a,1)/b<a$, a result that can be shown analytically \cite{CHEN1986}. These bounds can be tightened under certain circumstances using numerical methods \cite{LYON2021}.

The equation of the gamma distribution's inverse PDF (Eq. \ref{eq:gampdfinv}) has an interesting relationship to the median of the gamma distribution in the special case in which $a = 2$. For this special case, the median does have a known closed-form expression. In what follows, we will show the nature of this relationship.

Before proceeding, we also note that studies of the gamma distribution's median hold special interest, in part because of the median's connection to an equation of Srinivasa Ramanujan. Namely, the Laurent series expansion for the median of the gamma distribution can be derived from Ramanujan's theta function \cite{CHOI1994,RAMANUJAN1911}. The Ramanujan theta function has applications to many important problems in discrete mathematics and computer science, including hashing, caching, the birthday paradox, and resource contention \cite{FLAJOLET1995}. So progress on expressions for the median may benefit other areas of mathematics and computer science.

We begin by extending consideration of the distribution's median to the more general quantile function. The quantile function of a random variable, denoted $F^{-1}(p)$, is defined as the inverse of the random variable's CDF. For gamma distributions, observing that $F(x)$ is strictly monotonic, we can write

\begin{equation}
    F^{-1}(p)=x \phantom{-} \textrm{such that} \phantom{`} \phantom{`} F_X(x)=p
\end{equation}
The problem of finding the gamma distribution's median can now be reduced to deriving an expression for $F^{-1}(p)$ and then evaluating $F^{-1}(1/2)$. As with the CDF, the gamma distribution's quantile function has no known closed-form expression for all values of the shape parameter, $a$. For the special case of $\textrm{Gamma}(2,b)$ distribution, however, the CDF can be written as

\begin{equation}
\label{eq:gam2cdf}
F(x;a=2,b)=1-\frac{(b+x)e^{-x/b}}{b}
\end{equation}
Equation \ref{eq:gam2cdf} contains a product exponential. The quantile function of the $\textrm{Gamma}(2,b)$ can therefore be written in terms of the Lambert W function (see introduction)\cite{JIMENEZ2009}:

\begin{equation}
\label{eq:gam2quant}
F^{-1}(p;a=2,b)=-b\left(1+W_{-1}\left(\frac{p-1}{e}\right)\right)
\end{equation}
Equation \ref{eq:gam2quant}, bears a resemblance to Equation \ref{eq:gampdfinv}. Substituting $a=2$ and $k=-1$ into Equation \ref{eq:gampdfinv} yields

\begin{equation}
\label{eq:gam2pdfinv}
\left. 
f_{k}^{-1}(p;a=2,b)\right|_{k=-1}=-bW_{-1}\left(-b*p \right)
\end{equation}
Equations \ref{eq:gam2quant} and \ref{eq:gam2pdfinv} are both linear transformations of branch $k=-1$ Lambert W Functions. Resultantly, simple transformations of Equations \ref{eq:gam2quant} and \ref{eq:gam2pdfinv} can render an equality relationship. 

Recall that $f^{-1}(p)$ is multivalued for real inputs and real outputs. First, we scale the input probability by a factor of $1/(e\ast b)$. To isolate the greater of the two values outputted by  $f^{-1}(p)$ (i.e. the value corresponding to the $k=-1$ branch) we apply the max function to $f^{-1}(p)$. Lastly, we scale the max function output by $1/b$. Applying these transformations to $f^{-1}(p;a=2,b)$ gives

\begin{align}
\label{eq:gam2finvtranf}
 F^{-1}(1-p;a=2,b)/b+1 &= -b\left(1+W_{-1}\left(\frac{1-p-1}{e}\right)\right)/b+1 \nonumber
 \\ &=-W_{-1}\left(\frac{-p}{e} \right)
\end{align}

The requisite transformation of the gamma distribution quantile function for $a = 2$ requires that one substitute $1-p$ for $p$, scale the output of the quantile function by $1/b$, and then add $1$, which yields

\begin{align}
\label{eq:gam2quanttranf}
 F^{-1}(1-p;a=2,b)/b+1 &= -b\left(1+W_{-1}\left(\frac{1-p-1}{e}\right)\right)/b+1 \nonumber
 \\ &=-W_{-1}\left(\frac{-p}{e} \right)
\end{align}

The right-hand sides of Equations \ref{eq:gam2finvtranf} and \ref{eq:gam2quanttranf} are the same. Hence, 

\begin{equation}
\max\left[f_{k}^{-1}(p/(e*b);a=2,b)\right]/b=F^{-1}(1-p;a=2,b)/b+1
\end{equation}

Thus, in the special case in which the shape parameter $a = 2$, we have established that for all values of $p$—and hence for the value (i.e., $p = 1/2$) that corresponds to the median—a simple algebraic transform of the inverse density can be set equal to a simple algebraic transform of the quantile function.

%%       \cite{<label>} ==> [1]

%%




\bibliographystyle{elsarticle-num} 
\bibliography{main.bib}

\end{document}